%% file: main.tex
\RequirePackage{docswitch}
\pdfoutput=1
\setjournal{\flag}

\documentclass[twocolumn,nofootinbib]{\docclass}

\newcommand{\sfig}[2]{
	\includegraphics[width=#2]{#1}
}
\newcommand{\Sfig}[2]{
	\begin{figure}[thbp]
		\sfig{../figures/#1.pdf}{\columnwidth}
		\caption{{\small #2}}
		\label{fig:#1}
	\end{figure}
}

\newcommand{\Svwide}[2]{
	\begin{figure*}[thbp]
		\sfig{../figures/#1.pdf}{\textwidth}
		\caption{{\small #2}}
		\label{fig:#1}
	\end{figure*}
}

\newcommand{\rf}[1]{Figure \ref{fig:#1}}
\newcommand{\rsec}[1]{\S\ref{sec:#1}}
\newcommand{\rssec}[1]{\S\ref{subsec:#1}}
\newcommand{\ec}[1]{Eq.~(\ref{eq:#1})}

\newcommand\be{\begin{equation}}
	\newcommand\ee{\end{equation}}
\def\bea{\begin{eqnarray}}
	\def\eea{\end{eqnarray}}

\newcommand\full{the FCM}
\newcommand\gaussian{the GCM}
\newcommand\ctot{\mathcal{C}}

\usepackage{lsstdesc_macros}
\usepackage{footnote}

\usepackage{graphicx}
\graphicspath{{./}{./figures/}}
\begin{document}
	
	\title{Data Compression and Covariance Matrix Inspection: Cosmic Shear}
	\maketitlepre
	
	\begin{abstract}
		Covariance matrices are among the most difficult pieces of end-to-end cosmological analyses. In principle, for two-point functions, each component involves a four-point function, and the resulting covariance often has hundreds of thousands of elements. We investigate various compression mechanisms capable of vastly reducing the size of the covariance matrix in the context of cosmic shear statistics. 
		This helps identify which of its parts are most crucial to parameter estimation. We start with simple compression methods, by isolating and ``removing" 200 modes associated with the lowest eigenvalues, then those with the lowest signal-to-noise ratio, before moving on to more sophisticated schemes like compression at the tomographic level and, finally, with the Massively Optimized Parameter Estimation and Data compression (MOPED). We find that, while most of these approaches prove useful for a few parameters of interest, like $\Omega_m$, the simplest yield a loss of constraining power on the intrinsic alignment (IA) parameters as well as $S_8$. For the case considered --- cosmic shear from the first year of data from the Dark Energy Survey --- only MOPED was able to replicate the original constraints in the 16-parameter space. Finally, we apply a tolerance test to the elements of the compressed covariance matrix obtained with MOPED and confirm that the IA parameter $A_{\mathrm{IA}}$ is the most susceptible to inaccuracies in the covariance matrix.

	\end{abstract}
	
	% Keywords are ignored in the LSST DESC Note style:
	\dockeys{}
	
	\maketitlepost
	
	% ----------------------------------------------------------------------
	% {% raw %}
	
	\section{Introduction}
	\label{sec:introduction}
	Cosmic shear is a weak lensing effect caused by the large-scale structure of the universe and is an important tool for constraining cosmology. The most common way of obtaining information from cosmic shear is to use two-point functions and, as is often the case, this analysis assumes that the summary statistics have a gaussian distribution, thus requiring a covariance matrix. For a two-point data vector of length $N$, the covariance matrix is a symmetric $N\times N$ matrix with $N\times (N+1)/2$ individual elements that capture the auto and cross-correlation of the data vector. As the length of the data vector increases, the number of elements in the covariance matrix grows quadratically and becomes harder to evaluate. %One could potentially speed up computations and provide a simpler method of analyzing the covariance matrix by using compression schemes capable of significantly reducing the size of the matrix while still retaining relevant information about the parameters of interest.
	Compression schemes resolve this by significantly reducing the dimension of the matrix while still retaining relevant information about the parameters of interest, and also potentially speeding up computations. One way of accomplishing this is to use the Massively Optimized Parameter Estimation and Data compression (MOPED), in which, if the noise in the data does not depend on the model parameters, then the Fisher matrix for both the full and compressed covariance matrices coincides and the compression is said to be lossless \cite{Heavens:2000hjl, Tegmark:1997maa}. MOPED has been widely used in literature for a variety of applications, for example, CMB data \cite{Zablocki:2015zcm}, the redshift-space galaxy power spectrum and bispectrum \cite{Gualdi:2018mjl}, parameter-dependent covariance matrices \cite{Heavens:2017smv}, compression of the Planck 2015 temperature likelihood \cite{Heather:2019d}, weak lensing and galaxy clustering \cite{Ruggeri:2020rb}, and has been paired with a Gaussian Process emulator to analyze weak lensing data \cite{Mootoovaloo:2020}.
	
	We will focus on cosmic shear measurements from the Dark Energy Survey (DES)~\cite{Troxel:2017xyo} Year 1 release; the data vector has 227 elements, varying with angular separation and different pairs of tomographic redshift bins. Since our parameter space consists of 16 free parameters, we can use MOPED to reduce the $25,878$ independent elements of the covariance matrix, to only $136$.
	
	Apart from MOPED, we will be analyzing the covariance matrix with three other compression techniques: the first involves performing an eigenmode decomposition then discarding the modes associated with the lowest eigenvalues; the second approach removes those with the lowest signal-to-noise ratio. In order to obtain a compression competitive with MOPED in terms of shrinkage, i.e. about $10\%$ of the original size, we remove, in both cases, 200 such modes.
	
	Finally, the third method consists of a map-level compression \citep{Alonso:2017hhj}, where linear combinations of the tomographic maps are used to retain as much information as possible. Compression of the tomographic bin pairs then considerably reduces the size of the data vector of the two-point functions. For example, we will see that most of the information in the four tomographic bins used by DESY1 can be compressed into a single linear combination of those bins, or one Karhunen-Lo\'eve (KL) mode. Therefore, instead of $(4\times5)/2$ two-point functions for each angular bin, we need include only one or two.
	For this purpose, the data vector for each tomographic bin will have the same length, and so the angular cuts to the dataset and covariance matrix will be different from the ones used in the aforementioned DESY1 paper. The chosen covariance matrix has a dimension of $190 \times 190$. With one KL mode, we can compress the shear data vector down to $10\%$ of its original size, yielding 190 independent elements for the covariance matrix of the new data vector.
	
	In \rsec{methods}, we start by describing the dataset and the covariance matrices used. We then proceed to review each compression scheme and apply them to a DESY1 cosmic shear, demonstrating how well they reproduce the constraints obtained with the full covariance matrix.
	We follow this by showing that compression can be a useful tool to compare two different covariance matrices, in \rsec{comparison_matrices}.
	Our tolerance test is described in \rsec{tolerance}, where we investigate the change in parameter constraints resulting from the addition of noise separately to elements and eigenvalues of the covariance matrix. Finally, our conclusions are summarized in \rsec{conclusion}.
	
	% ----------------------------------------------------------------------
	\section{Methods}
	\label{sec:methods}
	
	\subsection{DES Cosmic Shear: Data and Analysis}
	\label{subsec:data_and_analysis}
	
	In this section, we introduce the data and covariance matrices that are used in this work. Our tests are carried out using cosmic shear statistics $\xi_\pm(\theta)$, focusing on the Year 1 results of the Dark Energy Survey \citep{Troxel:2017xyo,Abbott:2018cms} (DESY1). The data is divided into four tomographic redshift bins spanning the interval $0.20 < z < 1.30$, which yields 10 bin-pair combinations, each one containing 20 angular bins between 2.5 and 250 arcmin. We thus begin with 200 data points for statistic, giving 400 in total. We then apply the angular cuts described in \citep{Abbott:2018cms}, which removes the scales most sensitive to baryonic effects; this leaves 167 points for $\xi_+(\theta)$ and 60 for $\xi_-(\theta)$, resulting in 227 data points corresponding to the aforementioned $227 \times 227$ covariance matrix. 
	
	Table~\ref{tab:priors} shows the 16-parameters varied and the priors placed on them. Since cosmic shear is not sensitive to most of these, their constraints are largest dominated by the priors used. As such, throughout, we will only be showing constraints on three of those: the matter density parameter, $\Omega_m$, the amplitude of matter fluctuations, $S_8 \equiv \sigma_8 (\Omega_m/0.3)^{0.5}$, and the amplitude of the intrinsic alignment, $A_{\mathrm{IA}}$.
	
	To perform cosmological parameter inference we use the {\tt CosmoSIS} \citep{Zuntz:2015med, Lewis:2000taj, Kirk:2012mnras, Kilbinger:2009aa, Howlett:2012jcap, Bridle:2007njp, Takahashi:2012taj, Smith:2003mnras} pipeline, while employing the {\tt MultiNest} \citep{Feroz:2009fhb} sampler to explore the parameter space, with 1000 {\tt livepoints}, {\tt efficiency} set to 0.05, {\tt tolerance} to 0.1 and {\tt constant efficiency} set to True.
	
	\begin{table}
		\centering
		\begin{tabular} { l c} 
			\hline
			\hline
			Parameter		& Prior	\\ \hline
			Cosmological    & \\ [1ex]
			$\Omega_m$      & $\mathcal{U}(0.1, 0.9)$ \\
			$\log A_s$      & $\mathcal{U}(3.0, 3.1)$ \\
			$H_0 \mathrm{(km s^{-1} Mpc^{-1})}$	& $\mathcal{U}(55, 91)$\\
			$\Omega_b$      & $\mathcal{U}(0.03, 0.07)$ \\
			$\Omega_\nu h^2$& $\mathcal{U}(0.0005, 0.01)$ \\
			$n_s$           & $\mathcal{U}(0.87, 1.07)$ \\ [1ex]
			\hline
			Astrophysical       & \\ [1ex]
			$A_{\mathrm{IA}}$	& $\mathcal{U}(-5, 5)$ \\
			$\eta_{\mathrm{IA}}$& $\mathcal{U}(-5, 5)$ \\ [1ex]
			\hline
			Systematic      & \\ [1ex]
			$m^i$			& $\mathcal{G}(0.012, 0.023)$ \\
			$\Delta z^1$	& $\mathcal{G}(-0.001, 0.016)$ \\
			$\Delta z^2$	& $\mathcal{G}(-0.019, 0.013)$ \\
			$\Delta z^3$	& $\mathcal{G}(0.009, 0.011)$ \\
			$\Delta z^4$	& $\mathcal{G}(-0.018, 0.022)$ \\ [1ex]
			\hline
			\hline
		\end{tabular}
		\caption{List of the priors used in the analysis for parameter constraints ($\mathcal{U}$ denotes flat in the given range and $\mathcal{G}$ is gaussian with mean equal to its first argument and dispersion equal to its second). For the cosmological parameters, we fix $w = -1.0$, $\Omega_k =  0.0$ and $\tau =  0.08$. The astrophysical parameters are associated with the intrinsic alignment, they follow the relation $A_{\mathrm{IA}}(z) = A_{\mathrm{IA}}[(1+z)/1.62]^{\eta}$. Lastly, for systematics we have $m^i$ corresponding to the shear calibration and  $\Delta z^i$ for the source photo-$z$ shift, with $i = [1, 4]$ in both cases.}
		\label{tab:priors}
	\end{table}
	
	The covariance matrices are the following:
	\begin{itemize}
		\item the Full Covariance Matrix (FCM) used in the DESY1 analysis, which includes non-gaussian effects and super-sample variance; generated by {\tt Cosmolike} \citep{Krause:2016jvl};
		\item one containing only the gaussian part, which we will refer to as the Gaussian Covariance Matrix (GCM); generated by the same code used to analyze the KiDS-450 survey \citep{Kohlinger:2017sxk, Joachimi:2020blm}.
	\end{itemize}
	Thus, throughout, the covariance labels FCM and GCM differ for several reasons: first, they are two independent codes	and, second, although the code for the KiDS-450 survey does contain all the functionality in {\tt Cosmolike}, we ran \gaussian\ with the simplest settings in order to accentuate the differences. The ensuing discrepancies help us assess various validation techniques. Where not otherwise stated, the analysis and constraints will be performed on \full.
	
	\rf{Y1-constraints_wmS8A} shows the projected cosmological constraints for \full\ and \gaussian, using the same data vector and cuts. The 68\% CL constraints are as follows: for \full: $\Omega_m = 0.306^{+ 0.018}_{- 0.023}$, $S_8 = 0.784^{+ 0.054}_{- 0.06}$ and $A_{\mathrm{IA}} = 0.852^{+ 0.359}_{- 0.233}$; and for \gaussian: $\Omega_m = 0.309^{+ 0.017}_{- 0.023}$, $S_8 = 0.787^{+ 0.051}_{- 0.058}$ and $A_{\mathrm{IA}} = 0.948^{+ 0.329}_{- 0.22}$. This shows that the variations we introduced to the calculation of the two matrices are measurable in the parameter constraints.
	
	\Sfig{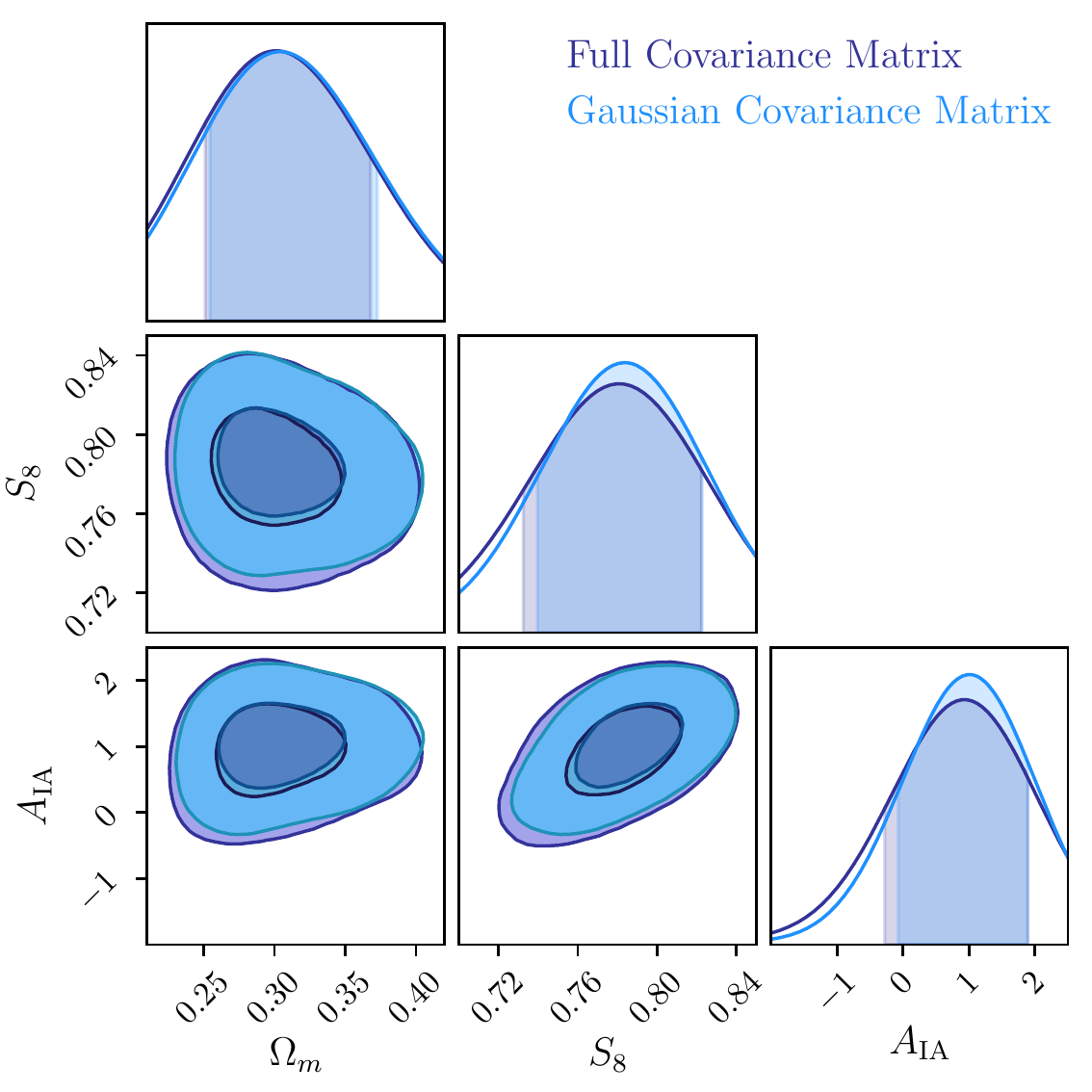}{Constraints on cosmological parameters $\Omega_m$ and $S_8$ and intrinsic alignment parameter $A_{\mathrm{IA}}$ for two covariance matrices produced for cosmic shear. The purple curve is for \full\ while the blue is for \gaussian. In the 16--dimensional parameter space, the volume of the posterior is about $22\%$ larger for the former.}
	
	% -------------------
	\subsection{Eigenvalues}
	\label{subsec:eigenvalues}
	
	Let us start with the easy task of analyzing the eigenvalues of the covariance matrix. Each eigenvalue is associated with a linear combination of the data vector, or a \emph{mode}.
	
	The idea is to remove the contribution of the lowest eigenvalues, since these are usually attributed to numerical noise and, as such, contain the least amount of information. The highest eigenvalues, on the other hand, are said to be the most informative \cite{Vogeley:1996} The procedure is simple, we first diagonalize the covariance matrix in order to calculate its eigenvalues then sort them into increasing order. Setting the lowest eigenvalues to zero would result in a non-positive definite (NPD) matrix, so we replace them instead with lower values (nine orders of magnitude lower), thus removing their effective contribution; we then transform back to the original basis and perform a cosmological analysis with the new covariance matrix.
	
	In order to reduce the covariance matrix to about 10\% of its original size, we follow the procedure above to discard the 200 eigenmodes with the lowest eigenvalues. The results reported in \rf{EigSNR-constraints_wmS8A} show a loss of constraining power on two of the three parameters shown. This is consistent with the fact that we are removing about 90\% of the information contained in the covariance matrices. However, it is inconsistent with the conjecture that the modes with lowest eigenvalues are irrelevant, in fact, constraints on $S_8$ for \full\ are $0.779^{+ 0.044}_{- 0.46}$, whereas, for the new covariance matrix, we obtain $0.725^{+ 0.076}_{- 0.083}$, showing an increase in the errors of almost 77\%. 
	It is then clear that this method is incompatible with a $10\%$ reduction, and so we must look for a different way of ordering the modes.
	
	\Sfig{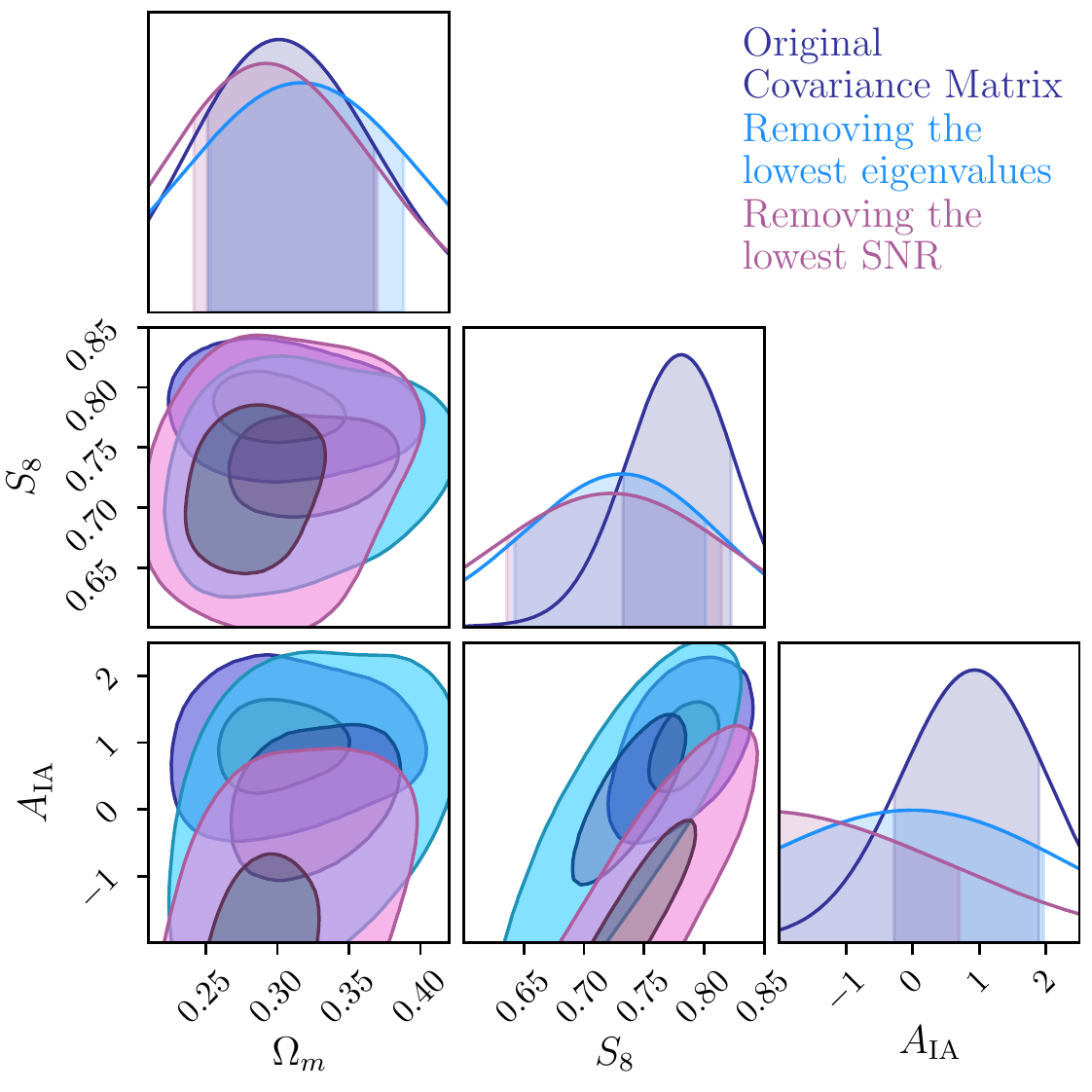}{Constraints on cosmological parameters $\Omega_m$, $S_8$ and the intrinsic alignment parameter $A_{\mathrm{IA}}$ for the original covariance matrix (in purple) and for the two new covariance matrices obtained in \rssec{eigenvalues} (in blue) and \rssec{snr} (in magenta).}
	
	\Svwide{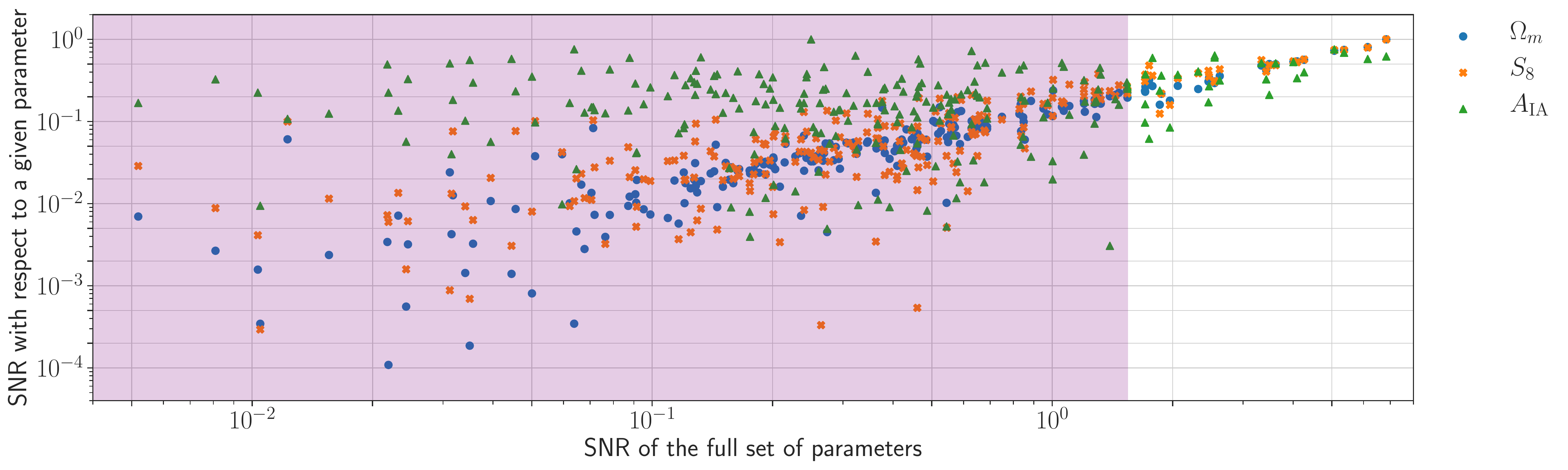}{Scatter plot for the relation between the signal to noise (SNR)  for each parameter (y-axis) against that for the full set of parameters (x-axis). The derivatives are shown with respect to $\Omega_m$ (blue circle), for $S_8$ (orange \textbf{x}) and for the intrinsic alignment parameter $A_{\mathrm{IA}}$ (green triangle). The purple rectangle spreads until the two hundred lowest values of SNR, which corresponds to the values that are modified for parameter constraints.}
	
	% -------------------
	\subsection{Signal-to-noise ratio}
	\label{subsec:snr}
	
	Instead of looking only at the ``noise'' -- or the eigenvalues of the covariance matrix -- a better way to assess the importance of modes is to consider the signal as well. We can define the expected signal-to-noise ratio (SNR) as
	\be
	\left(\frac{S}{N}\right)^2 = T_i C^{-1}_{ij} T_j\
	,\ee
	where $T_i$ is the predicted theoretical signal for the $i^{th}$ data point, given a fiducial cosmology, and $C$ is the covariance matrix. Repeated indices are summed in all cases, throughout this work. If $C$ were diagonal, then the eigenvectors would simply be the $T_i$s themselves, and not a linear combination of them, and we could estimate the SNR squared expected in each mode by just computing $T_i^2/C_{ii}$, with $ii$ denoting the diagonal element $i$. Then we could throw out the modes with the lowest SNR. Since this is not the case here, we have to first diagonalize $C$ and then order the values. We write the expected SNR squared as
	\bea
	\left(\frac{S}{N}\right)^2
	&=& \frac{v_i^2}{\lambda_i}\
	,\eea
	where $\lambda_i$ are the eigenvalues of the covariance matrix, which is diagonalized with the unitary matrix $U$, and the eigenvectors are 
	\be
	v_i\equiv U_{ij}^T T_j\
	,\ee
	with the superscript $T$ denoting the transpose. From a naive point of view, this makes it clear which modes should be kept and which should be dropped; modes $v_i$ for which $\left(v^2/\lambda\right)_i$ is small can be discarded. As we will later see, however, it is not as simple as that.
	
	After obtaining the SNR for the covariance matrix, we reduce the 200 lowest values to seven orders of magnitude lower, which is equivalent to increasing the noise (or decreasing the signal) of these modes. We then obtain a new covariance matrix with the corresponding modified SNR values. 
	The parameter constraints for this method are shown in \rf{EigSNR-constraints_wmS8A}, where we note that only $\Omega_m$ is well constrained (in agreement with those obtained with the original covariance matrix to within a $2\sigma$ interval). The constraining power on $A_{\mathrm{IA}}$ and $S_8$, on the other hand, is weakened, which suggests that the modes removed do indeed carry relevant information for these parameters.
	
	We can investigate this loss by tweaking our understanding of which modes carry information. The ``signal'' that these modes are ordered by is the amplitude of the data points.  The parameters, however, are sensitive to the shape as well as the amplitude.
	To address this, we can identify the SNR for each parameter individually by taking
	\bea
	\left(\frac{\partial S/\partial p_\alpha}{N}\right)^2 = \frac{(\partial v_i / \partial p_\alpha)^2}{\lambda_i}\
	,\eea
	where $\partial /\partial p_\alpha$ is the derivative with respect to each parameter. The importance of this procedure is illustrated in \rf{SNR_cuts200}, which shows the normalized SNR for a given mode on the $x$-axis against the SNR for $\Omega_m$, $S_8$ and $A_{\mathrm{IA}}$. The shaded region shows the 200 modes excluded in the previous analysis, where we see the presence of low SNR modes that contain information about the parameters. This is particularly true for the intrinsic alignment parameter $A_{\mathrm{IA}}$, which seems to explain the poor constraints shown in \rf{EigSNR-constraints_wmS8A}. As a result, simply cutting on raw SNR loses constraining power.
	
	On the other hand, as \citeauthor{Louca:2020} (2020) argues, removing the modes with the highest SNR is recommended in order to obtain a bias-free inference (another way would be to use a non-Gaussian likelihood). In light of that, we followed the same procedure used for removing the modes with the lowest SNR, but instead set the 200 highest modes to values several orders of magnitude lower. This yielded weaker constrains for not only for $S_8$ and $A_{\mathrm{IA}}$, but also for $\Omega_m$. We believe that this divergence was due to the large quantity of modes removed for our analysis and does not, in any way, invalidate the findings of the aforementioned work.
	
	%This analysis was also applied to the 200 modes with the highest SNR and yielded similar results. While \citeauthor{Louca:2020} (2020) argues that this would be recommended in order to obtain a bias-free inference (another way would be to use a non-Gaussian likelihood), we believe that our results diverge only due to the large quantity of modes that were removed for our purposes.
	
	% -------------------
	\subsection{Tomographic Compression}
	\label{subsec:tomographic_compression}
	
	The tomographic compression method of this section is based on a Karhunen-Lo\'eve (KL) decomposition for the shear power spectrum suggested by \citep{Alonso:2017hhj} and later applied to real space two-point function in \citep{Bellini:2019ssw} for the CFHTLens survey. Its implementation consists of finding the eigenmode --- in this case, a linear combination of the convergence in different tomographic bins --- with most of the signal-to-noise ratio contribution to the power spectrum, and then transforming the two-point function of this eigenmode into real space. This is not the most general compression method for the two-point function in real space, since the weight is dependent on the multipole $\ell$. However, as found in \citep{Bellini:2019ssw}, it is effective on the real space data, nonetheless.
	
	\begin{figure*}[thbp]
		\sfig{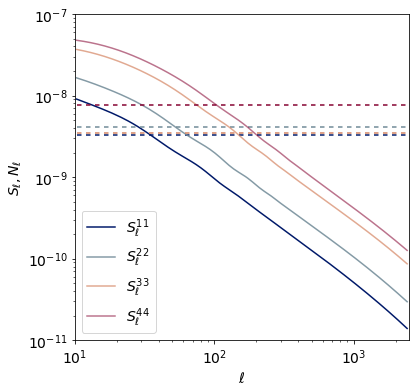}{0.8\columnwidth}
		\qquad \qquad \qquad
		\sfig{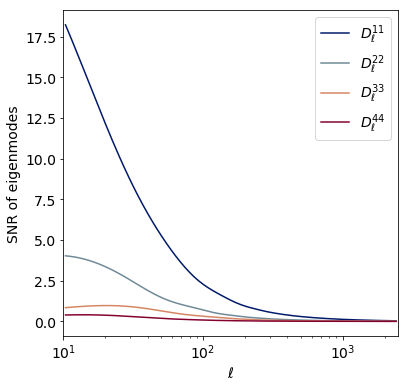}{0.8\columnwidth}
		\caption{\textbf{Left:} Shear power spectrum of \full. Solid lines are diagonal elements of the signal matrix $S_{\ell}$, and dashed lines are the diagonal elements of noise matrix $N_{\ell}$.
			\textbf{Right:} Signal-to-noise ratio matrix $D_\ell$ of the first to fourth KL-modes of the power spectrum on the left.  \label{fig:ClDl}}
	\end{figure*}
	
	Before diving into the derivation, it is worth summarizing the results. With {\tt CosmoSIS}, we calculate the shear angular power spectrum $\ctot_{\ell}$ of the convergence $\kappa^i$, where $i=[1,4]$ for the 4 tomographic bins probed by DES Year 1 with a fiducial cosmology at the best-fit parameters. We thus have $4\times 5/2=10$ pairs of bins for which we can compute spectra. The left plot in \rf{ClDl} shows the diagonal elements of the signal part, $S_\ell$, and of the noise part, $N_\ell$, of the spectrum. 
	The right-hand panel shows the signal to noise ratio for the KL-transformed eigenmodes, which we call $D_{\ell}$, ranging from $\ell = 10$ to $\ell = 2500$. That is, we identify a mode as $b_{\ell m} = r_i \kappa_{\ell m}^i$, where $r_i$ is the weight factor on the $i^{\text{th}}$ tomographic bins. We can see that the first KL mode contains most of the SNR contribution to the power spectrum. However, if we want to recover more information, we also should include the second and the cross mode between the first and second KL-mode.
	
	With the total power spectrum $\ctot_\ell = S_\ell+N_\ell$, we calculate the Karhunen-Lo\'eve (KL) modes for each $\ell$ (so we drop the $\ell$ subscript) via a general eigenvalue problem 
	\be
	\ctot e_p = \lambda_p N e_p
	.\ee
	The index $p$ in $e_p$ corresponds to the $p^{th}$ KL-mode of $\ctot$. Using Cholesky decomposition, $N = L L^T$ \footnote{Since we are dealing with real matrices, we replace the $\dagger$ with the transpose.}, we express the new observable as $b_p = e_p  L^{-1} \kappa$.
	And we find that $E_{\ell} = [e_1, e_2, \cdots] ^T$ is a basis transformation of basis so that the shear signal is diagonalized. We can now calculate the power spectrum $D_{\ell}$ for the new uncorrelated observable $b_{\ell m}$,
	\be
	D_{\ell}=\ \langle b_{\ell m} b_{\ell m}^T \rangle \ = E_{\ell} L^{-1} \ctot_{\ell} L^{-1} E^{T}_{\ell} = \Lambda_{\ell}\
	,\ee
	where $\Lambda_{\ell} = \text{diag}[\lambda_1, \lambda_2, \cdots]$ If we denote $E_{\ell} N^{-1}$ as $R_{\ell}$ and further write $U_{\ell}^{ij}=R^i_{\ell} R^j_{\ell}$, where $i$ and $j$ are the indices for the tomographic bin-pairs, we have the compression in terms of one simple linear combination,% of the $\ctot_{\ell}$,
	\be
	D_{\ell} = R_{\ell}^i \ctot_{\ell}^{ij} R_{\ell}^j = U_{\ell}^{ij} \ctot_{\ell}^{ij}\
	,\ee
	with $U_{\ell}^{ij}$ being the weight we will use to compress the two-point functions. We note that these KL-modes $b_{\ell m}^p$ are uncorrelated, so that their power spectrum $D_{\ell}^{pp'}$ is a diagonal matrix whose entries are 1+SNR of the corresponding eigenmodes. This allows us to compress ten tomographic bin-pairs to one, or two, by taking only the modes with the highest SNR.
	
	\begin{figure*}[thbp]
		\sfig{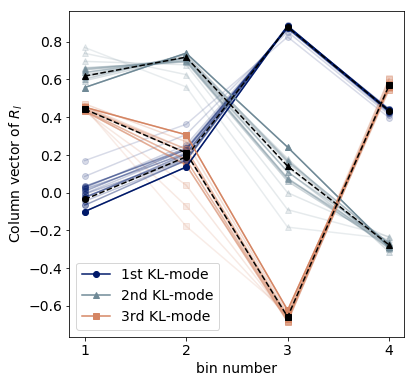}{0.8\columnwidth}
		\sfig{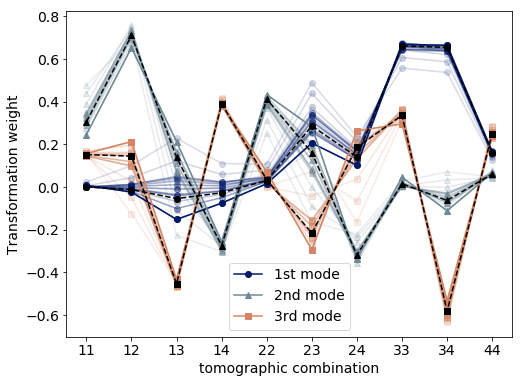}{1.02\columnwidth}
		\caption{\textbf{Left:} Column vectors of the matrix $R_{\ell}$, or $e^p_{\ell} N^{-\frac{1}{2}}$, for compressing the shear power spectrum $\ctot_{\ell}$. %, the changes in shades represent different $\ell$. 
			\textbf{Right:} Transformation on tomographic bin combination $U_{ij}$ constructed from the KL-eigenmodes. For both plots, the dashed black lines are the weighted average of each mode. The lightest shade represents $\ell = 10$ and the increment is $\Delta \ell = 10$ for each darker shade. \label{fig:kl-mode}}
	\end{figure*}
	
	We want, however, to eventually compress the two-point function data vector of DESY1, which is measured in the real space tomographic bin pair ${i, j}$ and related to the angular power spectrum $\ctot_{\ell}$ via
	\begin{equation*}
		\xi_{\pm}^{ij}(\theta) = \int \frac{\ell d \ell }{2\pi}J_{0/4}(\ell \theta) \ctot^{ij}(\ell)\ 
		.\end{equation*}
	In order to use linear combinations of all the tomographic bins, we need to ensure that the combination is $\ell$-independent, that is to say, the transformed two-point correlation function, $\Tilde{\xi}_{\pm}(\theta)$, can be directly calculated from other two-point functions. In fact, \rf{kl-mode} shows that the $U^{ij}(\ell)$ are generally $\ell$-independent, except for low $\ell$s, due to the existence of cosmic variance. Therefore, we have, 
	%In order to use only a linear combination of all the tomographic bins, we need to make sure that the combination is $\ell$-independent, that is to say, the two-point correlation function corresponding to $D_{\ell}$, $\Tilde{\xi}_{\pm}(\theta)$, can be directly calculated from other two-point functions. In \rf{kl-mode}, we show that compression matrices $U^{ij}(\ell)$ are generally $\ell$-independent, except for low $\ell$s, because of the existence of cosmic variance. Therefore, we have, 
	\bea
	\nonumber\Tilde{\xi}_{\pm}(\theta) &=& \int \frac{\ell d \ell }{2\pi}J_{0/4}(\ell \theta) D(\ell)\\\nonumber
	&=&\int \frac{\ell d \ell }{2\pi}J_{0/4}(\ell \theta) U^{ij}_\ell \ctot^{ij}(\ell)\\
	&=&\bar{U}^{ij}\xi_{\pm}^{ij}(\theta)\
	,\eea
	where $\bar{U}^{ij}$ is the average $U^{ij}_{\ell}$ given by,% weighted by the number of multipoles for each $\ell$ that is $2\ell+1$
	\be
	\bar{U}^{ij} = \frac{\int_{\ell _{\mathrm{min}}}^{\ell _{\mathrm{max}}} d\ell\, (2 \ell +1) U^{ij}_{\ell}}{\int_{\ell _{\mathrm{min}}}^{\ell _{\mathrm{max}}} d\ell\, (2 \ell +1)}\
	.\ee
	We make a more conservative angular cut than the one discussed in \cite{Troxel:2017xyo}, making sure that both $\xi_{\pm}(\theta)$ are uniform in regard to tomographic combinations. We consider an angular scale for  $\xi_+$ from $7.195'$ to $250.0'$, and for $\xi_-$ from $90.579'$ to $250.0'$. Therefore, for the purpose of exploring the KL-transform, the raw data vector has a length of 190. By shrinking 10 tomographic combinations for each angle into 1 KL-mode, the data vector is reduced to length 19, and so the number of elements in the covariance matrices has a compression of 99\%.
	
	In \rf{kl-mode}, we plot the normalized KL-eigenmode %$E_\ell^i$ of $\ctot_{\ell}$ 
	$e^p_{\ell} N^{-\frac{1}{2}}$ and its corresponding weight, $U^{ij}_\ell=R_\ell^i R_\ell^j$. %$U^{ij}_\ell=R_\ell^i R_\ell^j$. 
	Modes with increasing $\ell$ are plotted in increasing opacity of the color. While the KL-modes do vary by a slight amount for different $\ell$, their sensitivity to it is not very significant since they converge for higher $\ell$ to their weighted average, which we represent with the dashed black lines. %so we also take the weighted average of the eigenmodes $E_\ell^p$ and its quadratic form $U_\ell$ over $\ell$'s and plot them with black lines.
	For the first KL-mode, the tomographic bins with higher redshift are weighted more than those with low redshift. This is also shown in the right panel by the weight on tomographic combination that the combination of bin 3 and bin 4 carries most of the weight in the signal-to-noise ratio. This agrees with the fact that low-redshift galaxies are less affected by lensing than high-redshift galaxies, as indicated in the left panel of \rf{ClDl}.
	
	\begin{figure}[b]
		\includegraphics[width=\columnwidth]{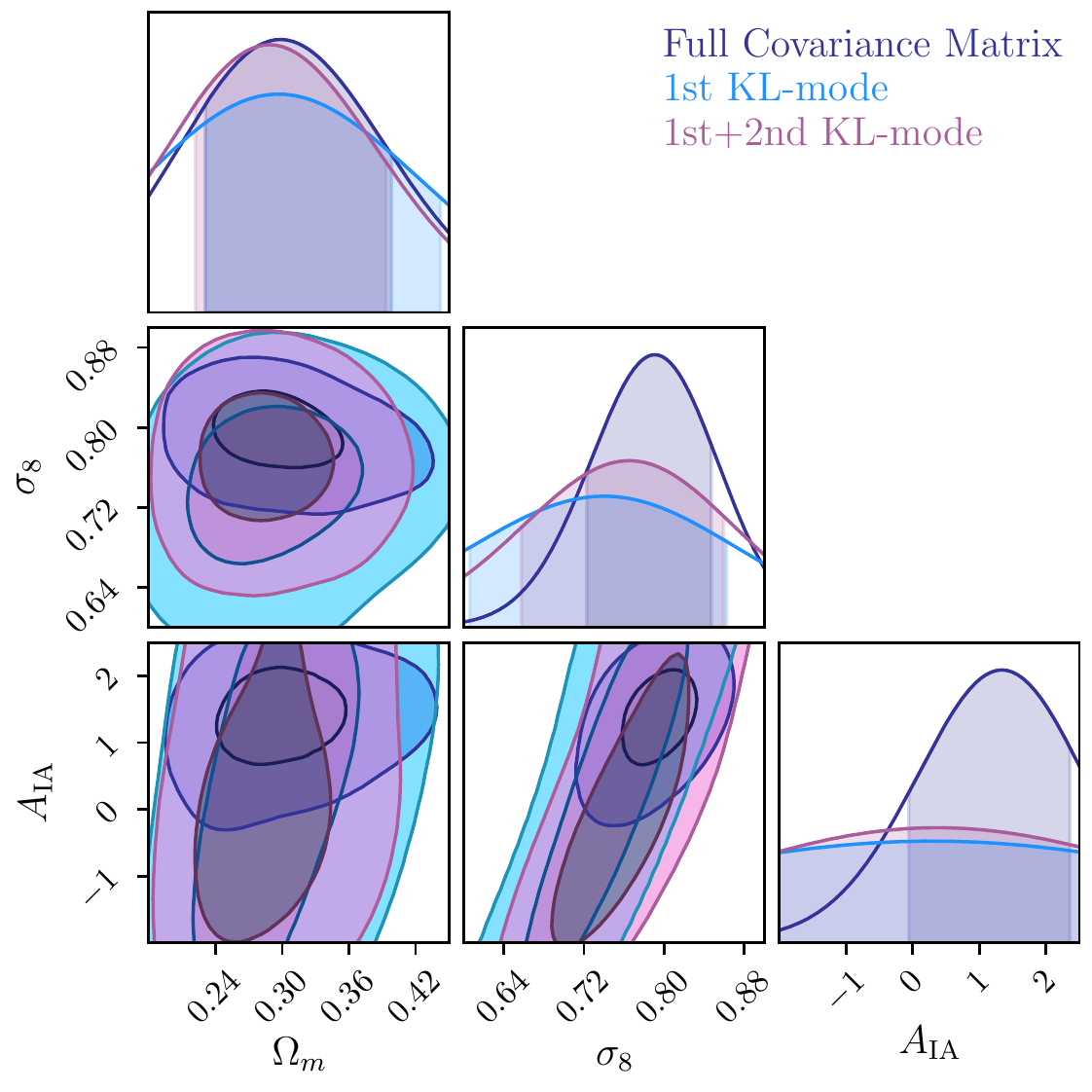}
		\caption{Cosmological constraints marginalized over all 16 parameters for the  $190 \times 190$ FCM and that compressed using the first KL-mode and the first two KL-modes.\label{fig:CompKL-constraints_wmS8A}}
	\end{figure}
	
	We ran the likelihood analysis as detailed in \rssec{data_and_analysis} with the first KL-mode and the first two KL-modes with their cross correlation mode, which correspond to a 10-to-1 and 10-to-3 compression, respectively, and show the parameter constraints on the $\Omega_m - S_8 - A_{\mathrm{IA}}$ plane in \rf{CompKL-constraints_wmS8A}. We do not include the third and fourth KL mode because they contain considerably less signal to noise. We can see that the first KL-mode is generally not sufficient to recover the information in the data vector. Since the first two modes contain most of the SNR contribution at a map level, we were able to recover the $\Omega_m$ constraints. However, information about the $S_8-A_{\mathrm{IA}}$ combination is clearly lost. This could be due to the fact that the SNR-prioritized modes are not the sensitive direction for these parameters, as was also the case in \rf{SNR_cuts200}. Indeed, the $S_8 - A_{\mathrm{IA}}$ plane shows a strong correlation between these two parameters. This likely explains why the constraints for $S_8$ widened: the KL-modes fail to break the degeneracy on $A_{\mathrm{IA}}$, which is mostly present in the modes that are insensitive to cosmic shear and are discarded in the compression process.
	
	% -------------------
	\subsection{Applying MOPED}
	\label{subsec:2pt_compression}
	
	\Svwide{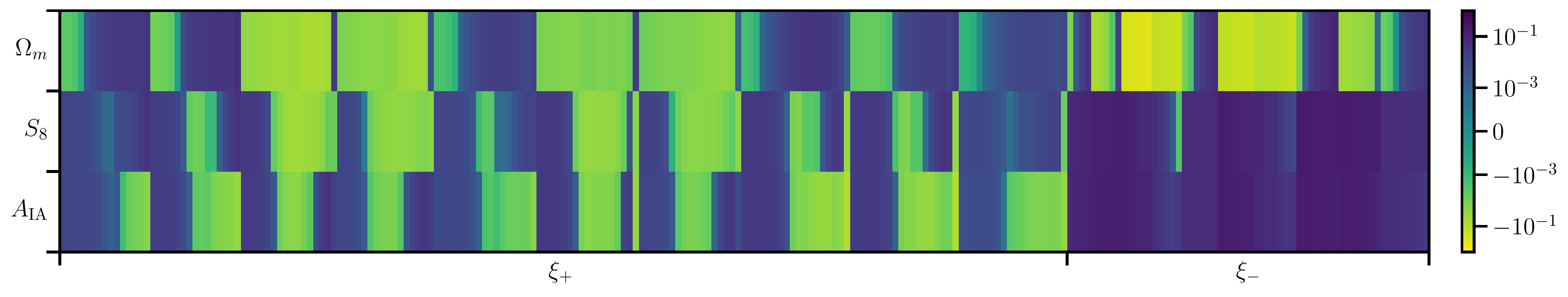}{An illustration of the 227 values of the weights corresponding to $\Omega_m$, $S_8$ and $A_{\mathrm{IA}}$ used for compressing the covariance matrices. Note the similarity of the weighting vectors for $S_8$ and $A_{\mathrm{IA}}$, and that the largest values correspond to the last 60 elements, i.e. those that we will use to compress the part of the covariance matrix that holds information for $\xi_-$.}
	
	The MOPED compression scheme takes place at the two-point level, with the compressed data vector containing linear combinations of the many two-point functions. In principle, this requires only $N_p$ linear combinations of the two-point functions where $N_p$ is the number of free parameters, and each mode, or linear combination, contains all the information necessary about the parameter of interest. 
	
	For each parameter $p_\alpha$ that is varied one captures a single linear mode
	\be
	y_\alpha = U_{\alpha i} D_i\
	,\ee
	where $D_i$ are the data points and the coefficients are defined as
	\be \label{eq:compression_scheme}
	U_{\alpha i} \equiv \frac{\partial T_j}{\partial p_\alpha} \, C^{-1}{}_{ji}\
	,\ee
	with $T_j$ being the theoretical prediction for the data point $D_j$ for a fiducial cosmology. An illustration of the matrix $U_{\alpha i}$ is shown in \rf{Weights_2pt}, showing the weighting vector for parameters $\Omega_m$, $S_8$ and $A_{\mathrm{IA}}$.
	
	The now much smaller data set $\{y_\alpha\}$, which contains $N_p$ data points, carries its own covariance matrix, from which $\chi^2$ can be computed for each point in parameter space. Propagating through shows that this covariance matrix is related to the original $C_{ij}$ via
	\be
	C_{\alpha\beta} = U_{\alpha i} C_{ij} U_{j\beta}\ 
	,\ee
	which also happens to be identical to the Fisher matrix of our likelihood. This compression was first suggested by \citeauthor{Tegmark:1997maa} (1997) for a single parameter only.  The non-trivial extension to multiple parameters, where the full Fisher matrix is reproduced with the compressed data, is the MOPED algorithm \citep{Heavens:2000hjl}. One difference here is that our weighing vector given by \ec{compression_scheme} does not carry the normalizing factor of Eq. (11) in \citep{Heavens:2000hjl}. In our case, the covariance matrix is  $227 \times 227$, while the number of parameters needed to specify the model is only 16, so $C_{\alpha\beta}$ is a $16\times 16$ matrix. We have apparently captured from the initial set of $(227 \times 228)/2 = 25,878$ independent elements of the covariance matrix a small subset (only 136) of linear combinations of these 26k elements that really matter. If two covariance matrices give the same set of $C_{\alpha\beta}$, it should not matter whether any of the other thousands of elements differ from one another.
	
	Ultimately, what matters is how well the likelihood does at extracting parameter constraints. Since most analyses assume a Gaussian likelihood, this boils down to how well the contours in parameter space agree when computing $\chi^2$ using two different covariance matrices.	
	
	\rf{Comp2pt-constraints_wmS8A} compares the constraints obtained for the compressed covariance matrix and data set with results from the full one. The two curves agree extremely well for the parameters shown: $\Omega_m$, $S_8$ and $A_{\mathrm{IA}}$. This is also true for all the other cosmological and intrinsic alignment parameters, where their mean values agree at the $1 \sigma$ confidence level. While the volume of the whole constrained parameter space does increase by about 13\%, the constraints for most parameters are less than 4\% broader, which shows that the information loss is negligible. 
	
	\Sfig{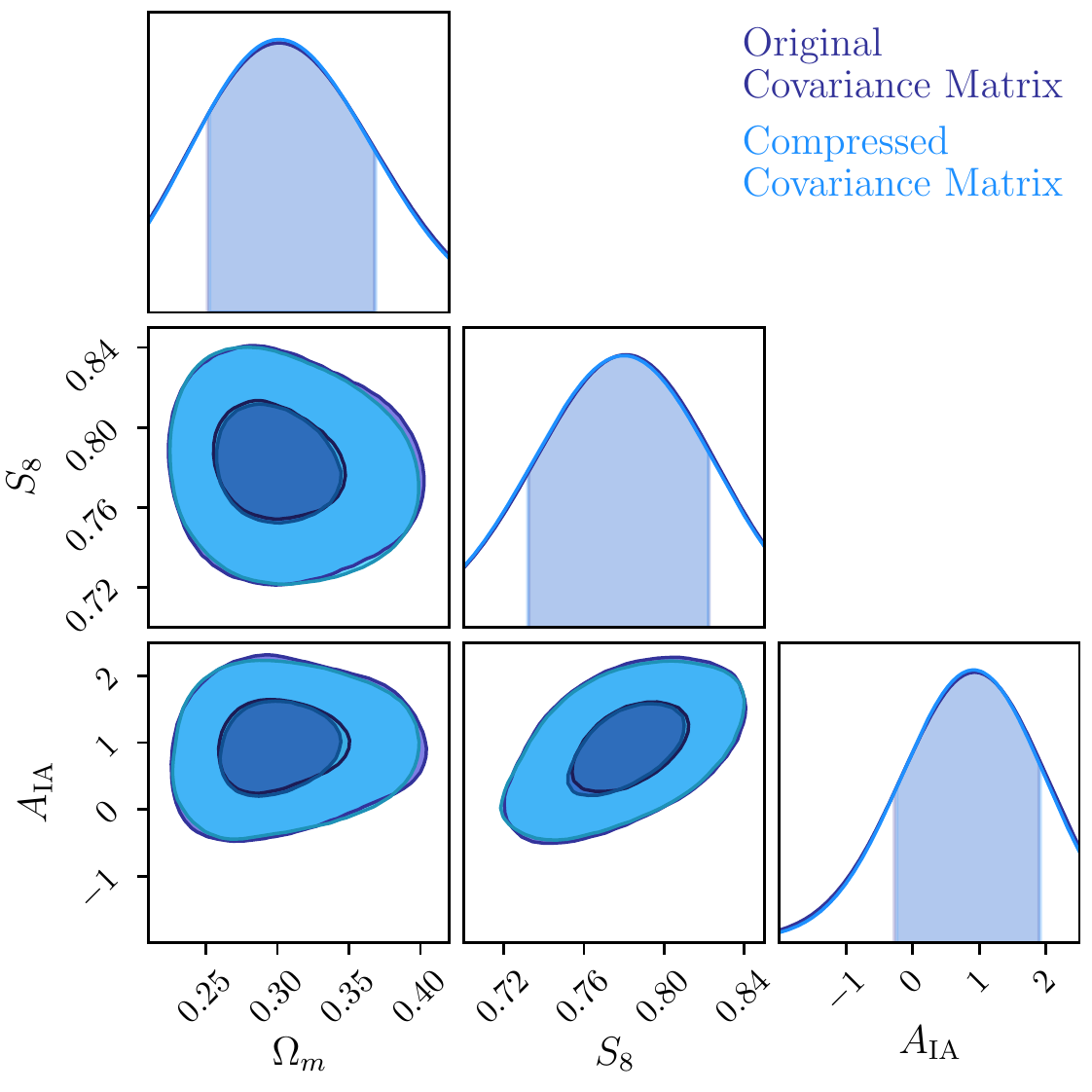}{Constraints on cosmological parameters $\Omega_m$ and $S_8$ and for the intrinsic alignment parameter $A_{\mathrm{IA}}$ for the original covariance matrix, FCM, (in purple) and for the compressed one (in blue).}
	
	% ----------------------------------------------------------------------
	\section{Comparison of Covariance Matrices}
	\label{sec:comparison_matrices}
	
	Armed with this information about compression, we now set out to compare the two covariance matrices, \gaussian\ and \full, described in \rssec{data_and_analysis}. 
	
	\subsection{Element-by-element comparison}
	\label{subsec:compare_one-one}
	
	We begin by performing an element-by-element comparison between the two covariance matrices. If there were only a single data point, then the covariance matrix would be one number and comparing two covariance matrices to try to understand why they give different constraints would be as simple as comparing these two numbers.  The simplest generalization is then to do an element-by-element comparison. We make a scatter plot of the elements of the two matrices in the bottom panel of \rf{Y1-scatter}, where we can see that the elements of \full\ are, in general, larger than \gaussian's, with many of the off-diagonal elements differing by 2 orders of magnitude or more.
	In some ways, this is useful and reassuring, as it aligns with what we see in the parameter constraints, in \rf{Y1-constraints_wmS8A}: larger elements in the covariance matrix translates to less constraining power. 
	
	The limitation of this method is that it remains unclear which of the differences are driving the final discrepancies in parameter constraints. This difficulty is an outgrowth of the increasing size of the data sets and hence the growing number of elements of the covariance matrix that any two codes are likely to disagree on. This element-by-element comparison, however, would prove much more useful if we fewer elements to compare. Towards that end, we turn to compressed covariance matrices.
	
	\Sfig{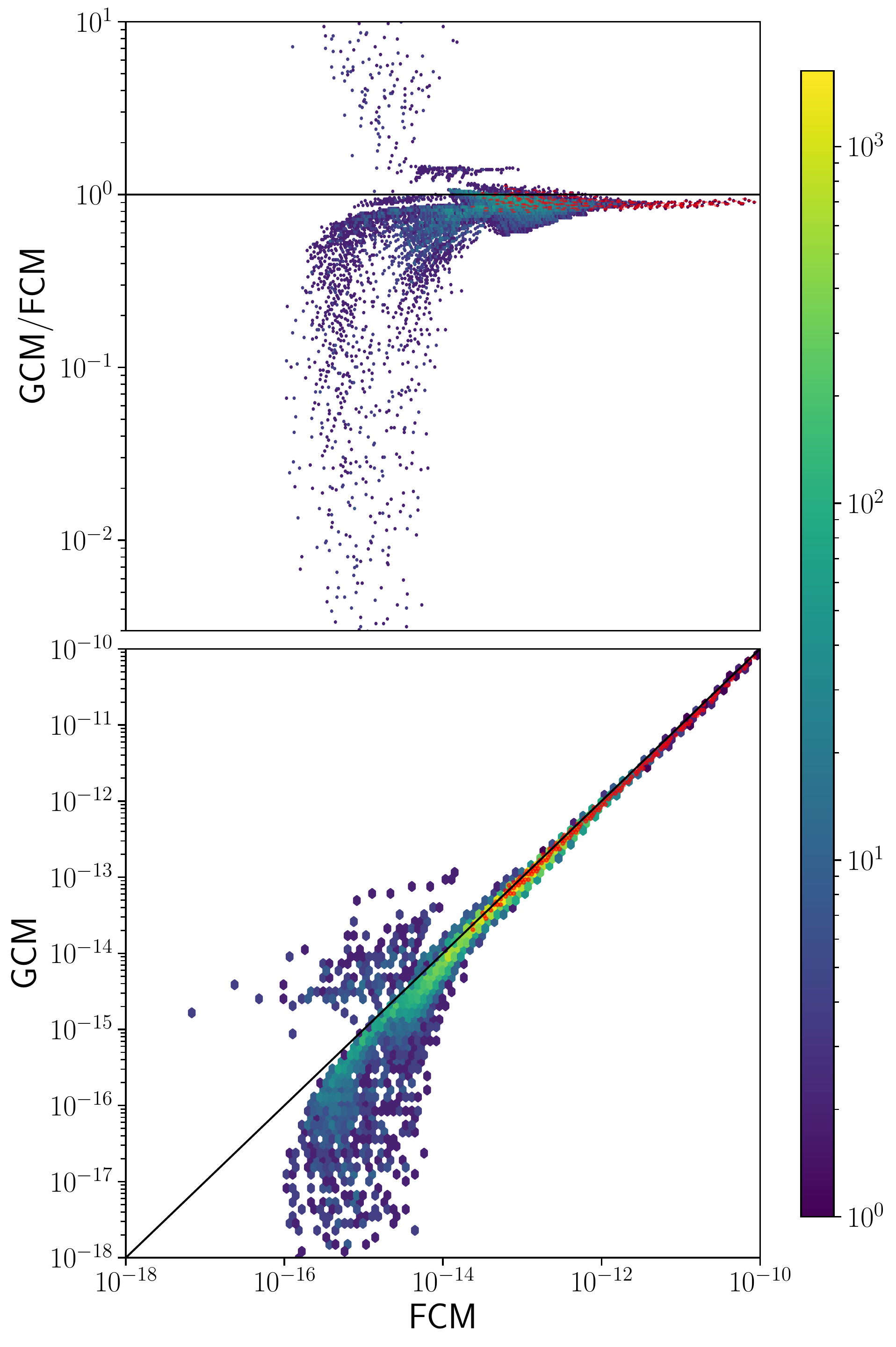}{In both plots, the red points refer to the diagonal elements, and the color bar varies according to the number of elements in one hexagonal bin, where the darkest blue color corresponds to only one element, and the brightest yellow shade to 2000. \textbf{Top:} Scatter plot of the ratio of the elements of \gaussian\ and \full\ vs \full\ value. For illustrative purposes, we draw a black, horizontal line at GCM/FCM = 1. \textbf{Bottom:} Density of the scatter plot of the positive elements of \gaussian\ and \full, with the black line showing FCM\ = GCM.}
	
	% -------------------
	\subsection{Compressed Matrices Comparison}
	\label{subsec:compare_compressed}
	
	\begin{figure}[b]
		\sfig{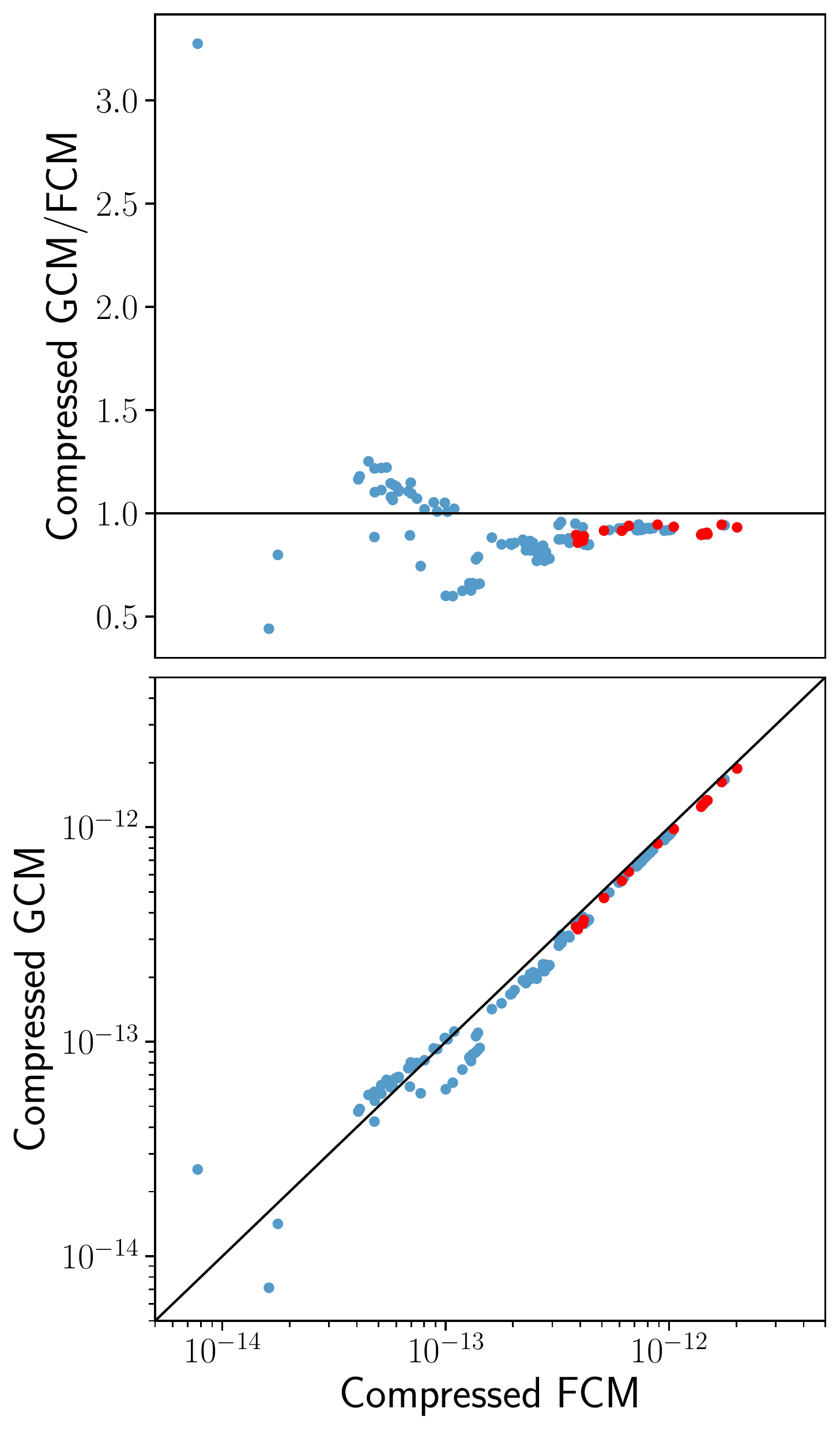}{0.85\columnwidth}
		\caption{{\small Results for the covariance matrices compressed following the procedure described in \rssec{2pt_compression}, with the red points corresponding to the diagonal elements.
				\textbf{Top:}  One-to-one scatter of the ratio of the elements of \gaussian\ and \full, over elements of \full. The black horizontal line is drawn at GCM/FCM\ = 1.
				\textbf{Bottom:}  One-to-one scatter of the elements of the compressed matrices, with the black line describing FCM\ = GCM.}}
		\label{fig:Comp2pt-scatter}
	\end{figure}
	
	Since we have shown that, out of all the compression schemes shown here, the only one capable of reproducing the original parameter constraints was MOPED, that is what we will be using in this section.
	
	We compress both covariance matrices using the same $U_{\alpha, i}$ (we also tried using different $U$'s for each and obtained similar results).
	%Here, we take two different approaches: first, we assume that $U_{\alpha, i}$ is the same for both covariance matrices and we calculate it with CL. The second approach is that each compression scheme must use the original covariance matrix that will be compressed, so that $U_{\alpha, i}$ will be different for each covariance matrix. We find that the mean values of the parameter constraints for the two methods both agree to $> 1\sigma$ of the original constraints. It is also crucial that the matrices used for comparison here are those obtained via the same compression scheme, so that we can be sure that their differences are indeed only related to the differences in the original matrices. 
	\rf{Comp2pt-scatter} show a one-to-one scatter plot of the compressed elements, which, as expected, exhibits a similar behavior to that observed in \rf{Y1-scatter}, with the elements of \full\ being larger than those of \gaussian. Here, however, the ratio of the diagonal elements is closer to 1, and the ratio of the diagonal elements goes up to only $\approx 2.3$. Perhaps even more importantly, there are much fewer points on this plot, since MOPED reduces the number of elements that need to be compared. These figures provide a greater insight into the relevant elements for parameter estimation: the dispersion is largely damped, and most of the elements are within $25\%$ of each other, which explains what we see in the parameter constraints. \rf{Comp2pt-correlation} shows the correlation matrix for \gaussian\ and \full, and the difference between the normalized off-diagonal elements. The small differences suggest that the root of the slightly looser constraints obtained with \gaussian\ is the larger diagonal elements of the MOPED-reduced covariance matrix. That is, a problem that initially required inspecting hundreds of thousands of elements is reduced to one involving only 16.
	
	\Sfig{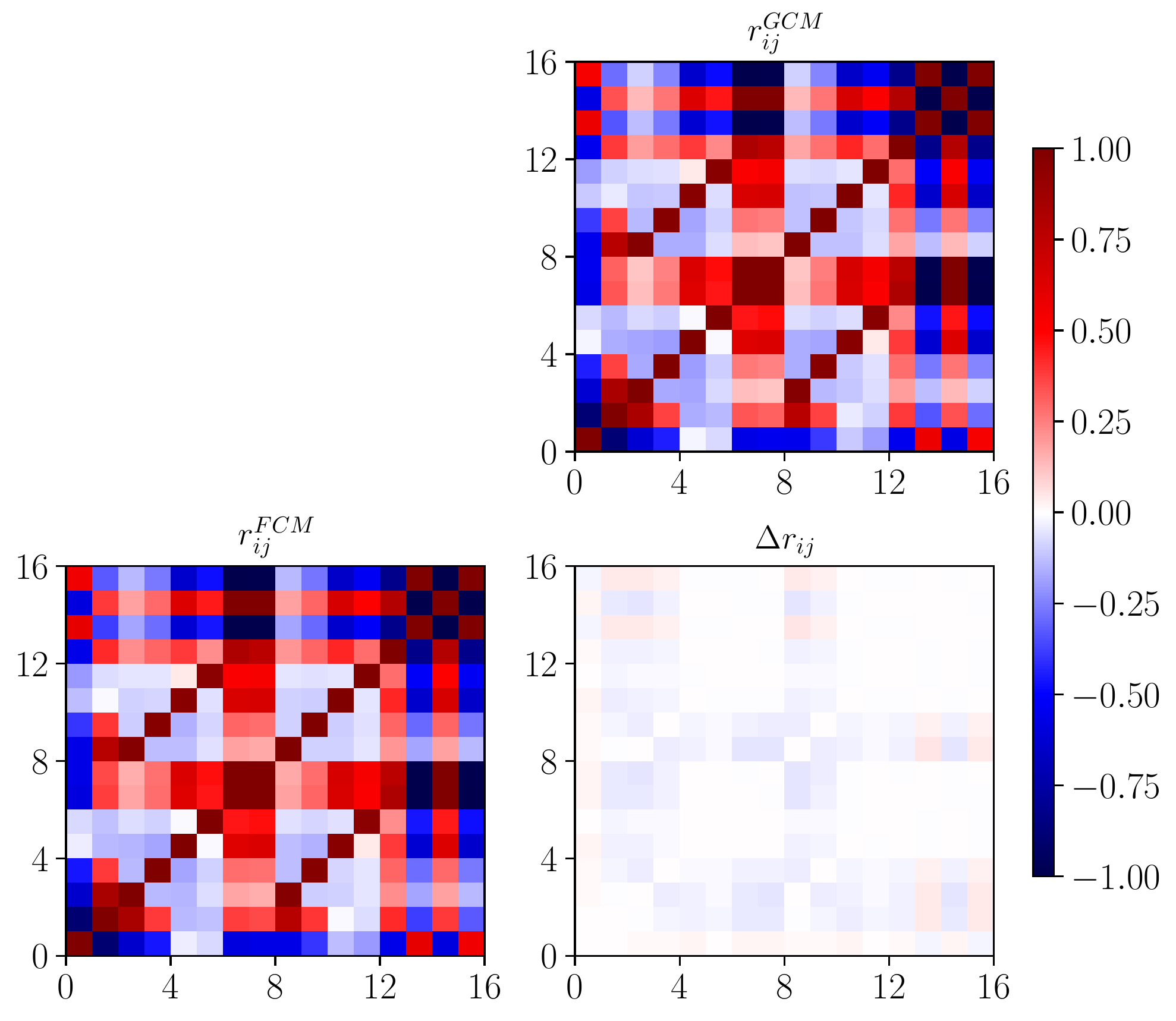}{The upper right and lower left plots display the correlation matrix for \gaussian\ and \full\ respectively, and the difference between them, $\Delta r_{ij}$, is shown on the lower right.}
	
	% ----------------------------------------------------------------------
	\section{Tolerance of the Compressed Matrices}
	\label{sec:tolerance}
	
	\Svwide{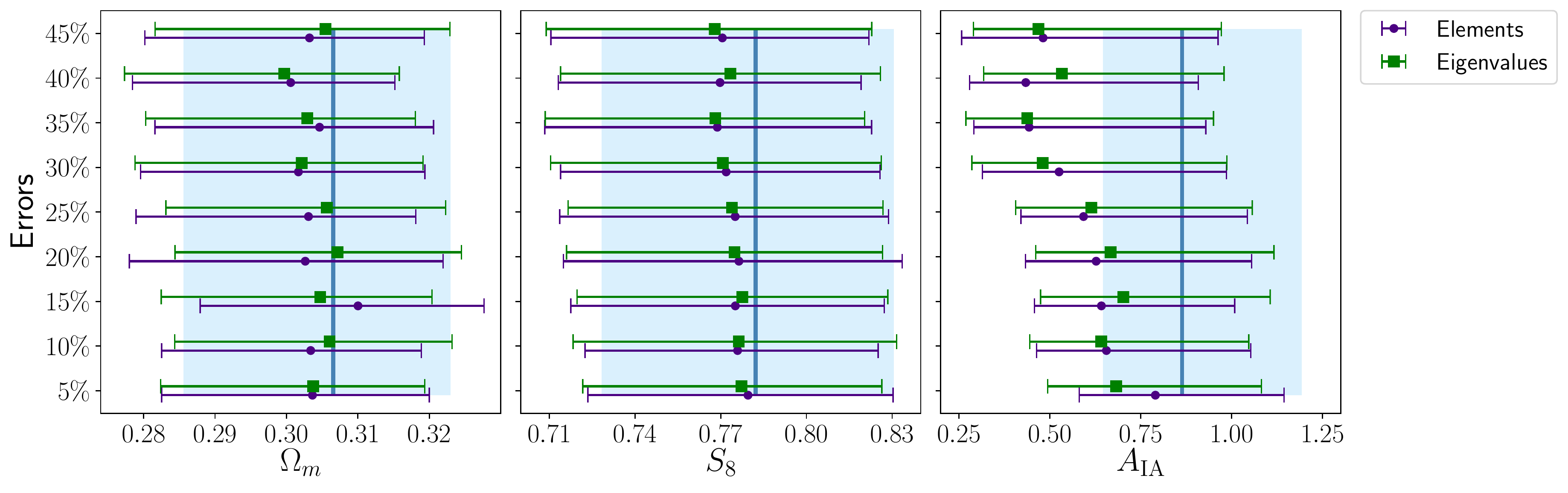}{An error plot showing the changes to the constraints for $\Omega_m$, $S_8$ and $A_{\mathrm{IA}}$ for errors added at $5\%, 10\%, 15\%, 25\%, 30\%, 35\%, 40\%$ and $45\%$ of the original elements (in purple) and eigenvalues (in green) of the compressed covariance matrix. The blue rectangle covers the 68\% CL interval obtained for \full, and the darker blue vertical line shows the mean value for the respective parameter.}
	
	Now that we have shown that we are indeed able to compress the covariance matrix into a much simpler and considerably smaller one, our next step is to analyze the amount of error the elements can tolerate while reproducing compatible parameter constraints.
	
	In the next two sections we test two different ways of perturbing the covariance matrix: first we consider an error to the elements themselves, then we follow a similar procedure to study the effects of introducing an error to the eigenvalues of the compressed covariance matrix. In both cases the perturbation is drawn in the following manner: consider that we want to test the impact of an error $x \%$; this can either be an increase of a decrease in the original element, or eigenvalue, as what we care about most is not whether the parameter constraints will be larger, but rather how different they are. For this error to be random, but centered at our desired percentage, we draw a $\delta$, for each new element/eigenvalue, from a Gaussian distribution, $\mathcal{G}(0,\frac{x}{100})$ and calculate the new value to be
	\be \label{eq:tolerance}
	C_{\alpha \beta}^{\mathrm{new}} = (1 + \delta)C_{\alpha \beta}^{\mathrm{old}}\ 
	,\ee
	where, for the eigenvalue, we replace $C_{\alpha \beta}$ with $\lambda_i$. This analysis is done only for \full, with errors ranging from $5 - 45 \%$, and for 50 realizations of the perturbed matrices.
	
	One of the concerns that arises when modifying the covariance matrix is that the resulting one has to be positive definite (PD). For this reason, in each section we also describe the steps taken to ensure this. Another intelligent way of guaranteeing PD would be to perturb the log of the covariance matrix. The issue, however, is how to introduce an error to the log matrix that would be similar to what we expect to see in the original covariance matrix. Introducing a $10\%$ error, for example, in such a matrix results in a perturbed covariance matrix with elements several orders of magnitude higher than the original one. A safer procedure would then be to perturb the log of its eigenvalues, but, since we have a section dedicated to perturbations to the eigenvalues themselves, we deemed this would be repetitive.
	
	% -------------------
	\subsection{Modifying the elements}
	
	Once we generate new values for each independent element, following \ec{tolerance}, we check for positive definiteness. Since the resulting matrix is, more often than not, not PD, we correct this by identifying the smallest negative eigenvalue and adding it to the diagonal \cite{Yuan:2008}. We check that, although doing this largely increases the values of the diagonal elements, less than $40 \%$ have a standard deviation of more than twice the original perturbation. 
	
	The constraints for $\Omega_m$, $S_8$ and $A_{\mathrm{IA}}$ are shown in \rf{Tolerance_constraints}, in purple, where the blue rectangle spans over the constraints for the unchanged compressed covariance matrix. The relative change in size for the 68\% CL interval is mostly $> 10 \%$ for the cosmological parameters; on the other hand, for the intrinsic alignment parameter $A$, the mean values are more than $1\sigma$ away from the original one and the loss in constraining power goes up to $\sim 30 \%$.
	
	% -------------------
	\subsection{Modifying the eigenvalues}
	
	Another way of introducing error to the covariance matrix is to perturb its eigenvalues. For a symmetric matrix, we have
	\be
	C = Q\Lambda Q^{-1}\ 
	,\ee
	where $\Lambda = \lambda I$, with $\lambda$ being the eigenvalues and $I$ the identity matrix; and $Q$ is a square matrix whose columns are composed of the eigenvectors of $C_{\alpha \beta}$. The eigenvalues are then perturbed as described in \ec{tolerance}, and the error, $\delta$ is drawn from $\mathcal{G}(0,\frac{x}{100})$, with the requirement that $|\delta| < 1$. We then have $\lambda^{\mathrm{new}} > 0$, and thus the perturbed covariance matrix associated with these new eigenvalues is PD. 
	
	The results for this method are also plotted in \rf{Tolerance_constraints}, in green. Despite the results following the same tendency as those of the last section, we find that about $80\%$ of the elements of the perturbed covariance matrices are within $10\%$ of their original value.
	
	% ----------------------------------------------------------------------
	\section{Conclusion}
	\label{sec:conclusion}
	
	In this work, we set out to explore different ways of compressing, comparing and analyzing covariance matrices, giving particular emphasis to the MOPED compression scheme. We started by looking at the parameter constraints of two $227 \times 227$ covariance matrices, \full\ and \gaussian, generated for DESY1 cosmic shear measurements, and saw that, although some of their elements differed by several orders of magnitude, they generated similar constraints. It was clear, then, that not all elements contribute equally to the parameter constraints, and we needed to employ increasingly complicated methods to try and locate the most relevant parts of the covariance matrix.
	
	The first step was then to analyze the eigenvalues. We began with the hypothesis that modes associated with the highest eigenvalues carry most information, as such, those with the lowest eigenvalues would contribute less to parameter estimation. Using this notion to compress the covariance matrix we ``removed" the lowest 200 eigenvalues, by setting them to several orders of magnitude lower. While the loss in constraining power for $\Omega_m$ was only around $20\%$, we saw a loss of about 77\% in the size of the constraints for $S_8$, and more than $100\%$ for $A_{\mathrm{IA}}$. Next, we moved on to the signal-to-noise ratio, and, using a similar procedure adopted for the eigenvalues, we ``removed" the modes with the lowest SNR. The results were similar to those obtained with the eigenvalue cuts and showed us that these modes did not contribute significantly to constraining some cosmological parameters, like $\Omega_m$, however constraints on the intrinsic alignment parameters, and even $S_8$ were more affected. This is consistent with the fact that the IA parameters are more sensitive to low SNR scales in cosmic shear, and it shows us that we need to look at the SNR per parameter before making any cuts, so that we do not lose important information for the parameters that we want to constrain.
	
	The next step was to shrink the covariance matrix by applying a tomographic compression, where we decompose the shear angular power spectrum into KL modes, then we look for modes with the highest SNR and compress shear data vector by the modes. We thus go from ten tomographic bin combinations to only one or two. The resulting covariance matrix, for one mode, is then reduced from $190 \times 190$ to $19 \times 19$ or $59 \times 59$, showing a reduction of about $99\%$ or $91\%$, respectively. We show, however, that one mode is not sufficient for constraining the parameters of our model, with the results being similar to our previous tests involving SNR: the constraints for $\Omega_m$, for example, are reproduced with the first and second KL-mode, but this is not the case for the IA parameters. Since essential information of IA parameters is contained in low SNR KL-mode, the high KL-modes failed to break the degeneracy of $A_{\mathrm{IA}}-S_8$ correlation, resulting in wider $S_8$ constraints. 
	
	Finally, we applied MOPED, which uses linear combinations of the data vector. By transforming the data vector and covariance matrix with a weighting vector that is parameter dependent, we were able to reduce the $227 \times 227$ matrix to a $16 \times 16$ matrix.%, and since the Fisher matrix is identical for both the original and compressed ones, the compression scheme is lossless. 
	We show that the cosmological analysis using this compressed matrix reproduced similar constraints to the DESY1 analysis, for an uncompressed covariance matrix. We also showed a comparison of the elements of the compressed covariance matrix for \full\ and \gaussian\ and found that the new elements show reasonable agreement, with their correlation matrices being very similar, and the diagonal elements showing a percentage difference of less than $15\%$.
	
	Given these results, we successfully show that MOPED is the only compression scheme, out of the ones considered in this work, capable of capturing all the relevant information required to reproduce reliable parameter constraints for the 16 parameters of interest. It is worth noting here that compression does not automatically speed up the computation for parameter inference, since it has to be redone for every point in the parameter space. Recent work has been done by \cite{Mootoovaloo:2020} to address this problem by using Gaussian Processes to generate the compressed theory.
	
	When looking at the one-to-one element comparison of \full and \gaussian, in \rf{Y1-scatter}, the region of large variance suggests that there could be considerable differences in the parameter constraints. We see, however, in \rf{Y1-constraints_wmS8A}, that this is not the case. This becomes clearer when comparing the elements of the compressed covariance matrices, where, while they do follow the same tendency as the full comparison, only a smaller portion of the elements display a greater dispersion.
	
	One last step was taken to analyze the error tolerance of the compressed FCM. We adopted two ways of doing this, by introducing error taken from a Gaussian distribution for $5 - 45 \%$ of the original 1) element and, 2) eigenvalue of the compressed covariance matrix. For the latter, we checked that only about $20 \%$ of elements of the resulting, perturbed, covariance matrix showed errors within the expected value, while the vast majority had only about a $10\%$ error. In both cases, however, the results were similar: for the cosmological parameters $\Omega_m$ and $S_8$, the $2\sigma$ constraints changed by about $7\%$, on average, while for the intrinsic alignment parameter $A_{\mathrm{IA}}$, the constraints were up to $30\%$ larger. Finally, we highlight the increasing shift, in the mean values of $A_{\mathrm{IA}}$, to about $32\%$ smaller than those obtained with the uncompressed FCM; while for the cosmological parameters this was only about $5\%$, in general.

	% ----------------------------------------------------------------------
	% ----------------------------------------------------------------------
	\subsection*{Acknowledgments}
	
	The authors wish to thank Sukhdeep Singh, Hung-jin Huang, Patricia Larsen, Benjamin Joachimi, Mike Jarvis and Rossana Ruggeri for useful discussions. %We also thank our esteemed internal reviewers from the LSST Dark Energy Science Collaboration: David Alonso, Jonathan Blazek, Alan F. Heavens and Patricia Larsen.
	This paper has undergone internal review by the LSST Dark Energy Science Collaboration. The internal reviewers were David Alonso, Jonathan Blazek and Alan F. Heavens.
	
	%%% Here is where you should add your specific acknowledgments, remembering that some standard thanks will be added via the \code{desc-tex/ack/*.tex} and \code{contributions.tex} files.
	
	\input{contributions}
	
	% Standard papers only: author contribution statements.
	
	% This work used TBD kindly provided by Not-A-DESC Member and benefitted from comments by Another Non-DESC person.
	
	% Standard papers only: A.B.C. acknowledges support from grant 1234 from ...
	
	\input{desc-tex/ack/standard} % also available: key standard_short
	
	% ----------------------------------------------------------------------
	% ----------------------------------------------------------------------

	% This work used some telescope which is operated/funded by some agency or consortium or foundation ...
	
	% We acknowledge the use of An-External-Tool-like-NED-or-ADS.
	
	%{\it Facilities:} \facility{LSST}
	
	% Include both collaboration papers and external citations:
	\bibliography{main}

\end{document}

%% file: contributions.tex
	%T.F. and T.Z. contributed equally on writing the main paper as well as implementing the covariance comparison and compression. N.C. contributed to the compression code. All authors participated in the discussion and gave valuable suggestions.
	
	The contributions are listed below. T.F. contributed to the manuscript, led the analysis for the eigenvalues, SNR and MOPED, as well as the comparison of the compressed covariance matrices. T.Z. contributed to the manuscript, participated and contributed substantially to all analysis, and led the analysis for the tomographic compression. N.C. contributed to the compression code. S.D. proposed the project, the analyses, led the discussions and also contributed to writing and editing the manuscript.

%% file: desc-tex/ack/standard.tex
The DESC acknowledges ongoing support from the Institut National de Physique Nucl\'eaire et de Physique des Particules in France; the Science \& Technology Facilities Council in the United Kingdom; and the Department of Energy, the National Science Foundation, and the LSST Corporation in the United States.  DESC uses resources of the IN2P3 Computing Center (CC-IN2P3--Lyon/Villeurbanne - France) funded by the Centre National de la Recherche Scientifique; the National Energy Research Scientific Computing Center, a DOE Office of Science User Facility supported by the Office of Science of the U.S.\ Department of Energy under Contract No.\ DE-AC02-05CH11231; STFC DiRAC HPC Facilities, funded by UK BIS National E-infrastructure capital grants; and the UK particle physics grid, supported by the GridPP Collaboration.  This work was performed in part under DOE Contract DE-AC02-76SF00515. T.F also acknowledges financial support from CAPES and FAPES.